\documentclass[pre,twocolumn,showpacs]{revtex4}

\usepackage{amsmath}
\usepackage{amssymb}
\usepackage{graphicx}

\begin{document}
\narrowtext
\title{Minority game with local interactions due to the presence of herding behavior}

\author{Daniel Oliveira Cajueiro and Reinaldo Soares De Camargo}
 \affiliation{Department of Economics, Catholic University of Brasilia, 70790-160,
Brasilia, DF, Brazil.}
%\email{danoc@ucb.br}
\begin{abstract}
In this paper, we introduce a framework to study local
interactions due to the presence of herding behavior in a minority
game. The idea behind this approach is to consider that some of
the agents who play the game believe that some of their neighbors
are more informed than themselves. Thus, in this way, these agents
imitate their most informed neighbors. The notion of neighborhood
here is given by a regular network, a random network or a small
world network. We show that under herding behavior the cooperation
between the agents is less efficient than that one which arises in
the standard minority game. On the other hand, depending on the
topology of the network, we show that that the well known curve
volatility versus memory, which caracterizes the minority game, is
a monotone decreasing curve.

Keywords: Complex networks, econophysics, herding behavior,
minority games, phase transition.

\end{abstract}

\pacs{ 02.50.Le, 05.65.+b, 05.70.Fh, 87.23.Ge, 89.65.Gh,
89.75.-Hc}

 \maketitle

%\emph{Lattices with $K$ neighbors.}
%\begin{figure}
%\includegraphics[width=7cm,height=7cm]{redereg.eps}
%\caption{A typical regular lattice that arises with $n=20$ and
%$K=8$.} \label{fig:redereg}
%\end{figure}
\section{Introduction}
In these last years, one of the most interesting contributions of
the statistical physics to the social sciences has been to study
the dynamics and the collective behavior of populations of agents
who compete for limited resources. In particular, the so-called
minority game (MG) introduced in~\cite{chazha97} as a
simplification of the Arthur's El Farol Bar~\cite{art94}
attendance problem is one of the simplest complex systems that
belong to this class. This game can be described in the following
way. At a given instant of time, an agent who belongs to the
population chooses between two opposing actions namely $a=\pm 1$
\footnote{In a financial market, for instance, this means to buy
or to sell an asset.}. Since the resources are limited, the
objective of each agent is to choose the side shared by the
minority of the population. The difficulty is that each agent does
not know what the others will choose. The agent decides his/her
next action based only on a global information, which is the
sequence of the last $M$ outcomes of the game, where $M$ is said
to be the memory of the agents. Therefore, there is no best
solution for the problem, i.e., the agents do not know what is the
best strategy \footnote{A strategy defines which action to take in
each state.} to deal with the game. Since there are only two
possible choices, the number of states is $2^M$ and there are at
all $2^{2^M}$ strategies. In~\cite{chazha97}, each agent has a
fixed number of strategies that do not change over time. Since
agents have different beliefs, the strategies differ from agent to
agent. At every turn of the game, the agents use their strategies
with highest scores\footnote{The strategies with highest scores
are those which were successful in the previous turns of the
game.}.

This standard MG presented above has been very well-studied -- a
revision of these attempts may be found for instance
in~\cite{johjef03,coo05,chamar05}. One of the most surprising
properties presented first in~\cite{savman99} is that if one plots
the ratio $\sigma^2/N$ as a function of $\alpha=2^M/N$, one may
conclude: (1) For small values of $\alpha=2^M/N$, the agents would
perform worse than if they had taken purely random decisions. (2)
For large values of $\alpha=2^M/N$, the agents' performance
approaches the random decision. (3) There is a critical value of
$\alpha=\alpha_c$ where the resources of the game are used in the
best way possible, i.e., the ratio $\sigma^2/N$ is the minimal
possible -- which suggests a non-equilibrium phase transition from
the so-called low-$M$ phase to the high-$M$ phase. The low-$M$
phase is characterized by a decrease in $\sigma^2/N$ as
$\alpha=2^M/N$ increases and the high-$M$ phase is characterized
by a increase in $\sigma^2/N$ as $\alpha=2^M/N$ increases. (4) The
behavior of the MG does not depend on the number of strategies
available for each agent.

In this paper, we introduce a version of the standard MG with
local interactions and exchange of local information. Actually,
this is not the first paper that presents the exchange of local
information and local interactions in some version of the MG. As
far as we know, the first attempt in this line was presented
in~\cite{pacbas00} where was developed a version of the Kauffman
network using some rules of the minority game and each agent
receives input from a fixed number of agents in the system. Other
formulations of MGs with local interactions may be found
in~\cite{kalsch00,moerio02,galler02,quawan03,chacho04,lisav04,burcev04,carcev04,shawan05,kir05}.
In our paper, differently of the others cited above, the exchange
of local interaction emerge from the imitation of the most
informed agents, i.e., some of the agents who play the game
believe that some of their neighbors are more informed then
themselves. This phenomenon called herding behavior happens when
an agent blindly follows the decision of other agent. The economic
theory says that it is rational even when the former agent's
signals suggest a different decision and it is
ignored\footnote{For details, see, for instance,~\cite{ban92}.}.
In the case here, an agent follows another agent strategy, if
he/she believes that there is an agent more informed than
hinself/herself, i.e., this most informed agent is more likely to
know the decision of the minority. Actually, one should notice
that if not one but many agents follow the most informed agent of
their neighborhood, then, in the future, the most informed agent
will be in the majority and he will no be followed anymore. In
this paper, we investigate how this kind of policy performed by
the agents affect the dynamics of the standard MG.

 \section{The MG with herding behavior} The game considered here has a framework quite
similar to the standard one presented in~\cite{chazha97}. The
difference is described bellow. In each time step, each agent
looks for the most informed agent located in his/her neighborhood.
The most informed agent here is the one that has the highest
scored strategy. Then, each agent compares his/her highest scored
strategy with the highest scored strategy of his/her most informed
neighbor. If the agent's highest scored strategy is higher scored
than the highest scored strategy of the most informed agent in
his/her neighborhood, then he/she follows his/her own strategy.
Otherwise, he/she follows his/her most informed neighbor highest
strategy.

\section{The notion of neighborhood}. The notion of neighborhood is
provided by one of the following networks: (1) a regular network;
(2) a random network~\cite{erdren60} or (3) a small world
network~\cite{watstr98,wat99}. While the regular network and the
random network are chosen to be used as references, the small
world network is chosen since it presents a topology that is
likely to happen in real situations of social
interaction~\cite{caj05}. Then, using one of these network
structures, each agent of the minority game is located in a node
of the network.

\section{Results.} 
\begin{figure}
\includegraphics[width=7cm,height=7cm]{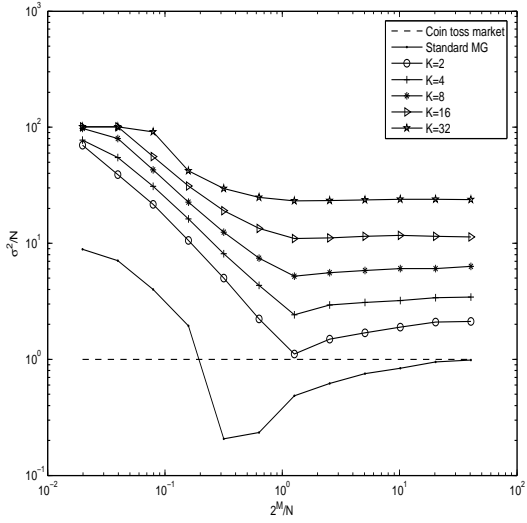}
\caption{The ratio $\sigma^2/N$ as a function of $\alpha=2^M/N$.
We compare here the standard MG with the modified MG presented in
this paper using regular networks.} \label{graf1}
\end{figure}

\begin{figure}
\includegraphics[width=7cm,height=7cm]{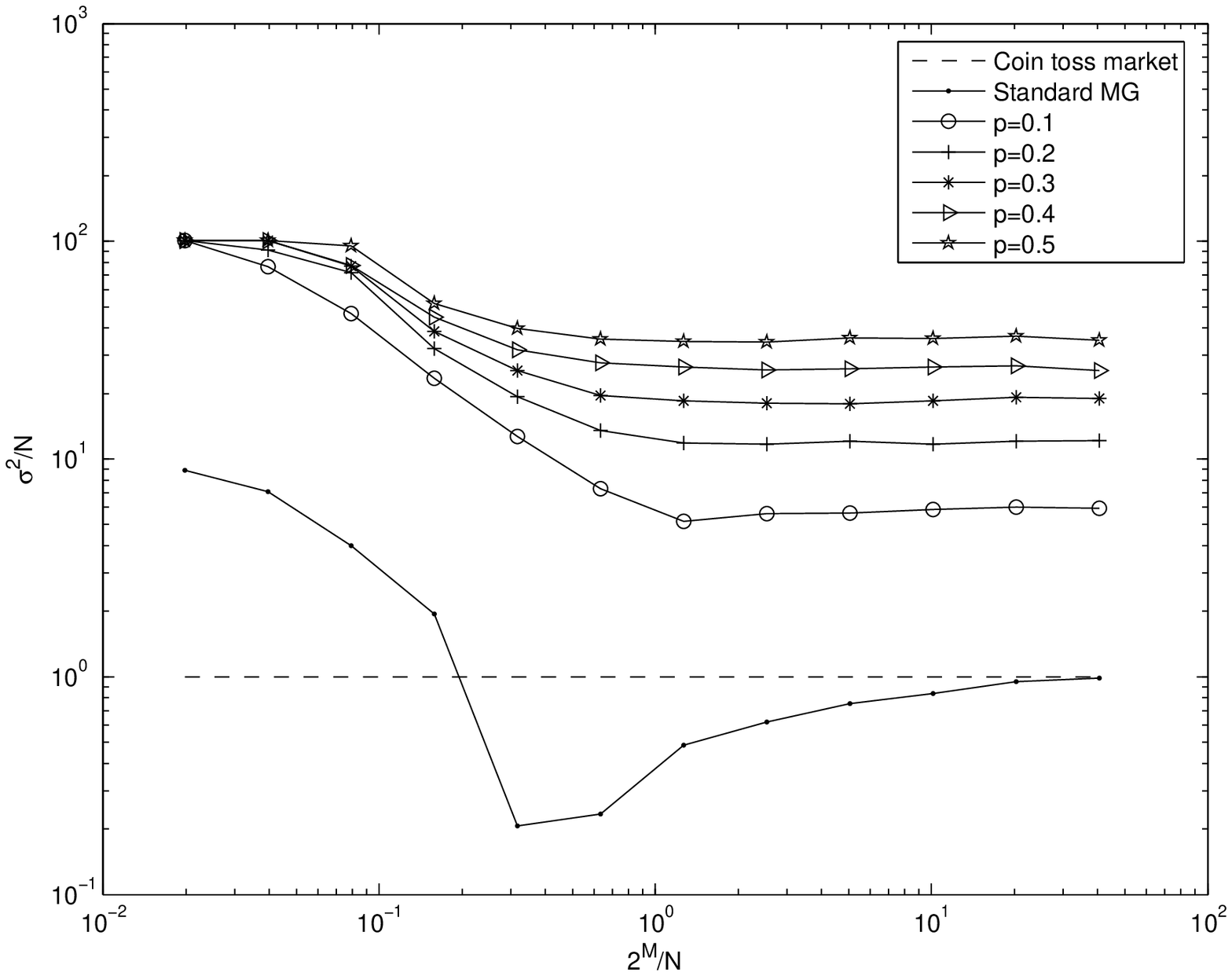}
\caption{The ratio $\sigma^2/N$ as a function of $\alpha=2^M/N$.
We compare here the standard MG with the modified MG presented in
this paper using random networks.} \label{graf2}
\end{figure}

\begin{figure}
\includegraphics[width=7cm,height=7cm]{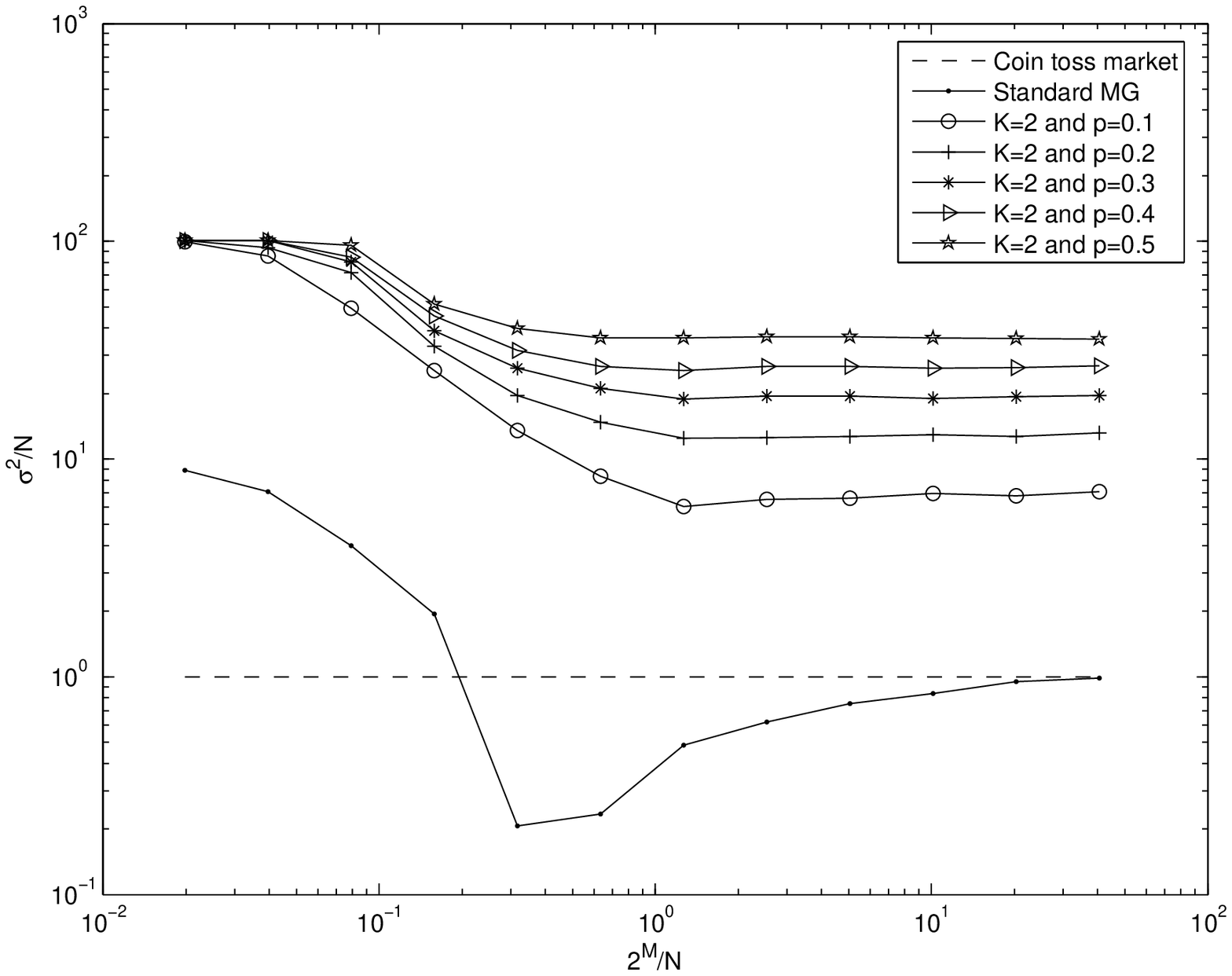}
\caption{The ratio $\sigma^2/N$ as a function of $\alpha=2^M/N$.
We compare here the standard MG with the modified MG presented in
this paper using small world networks with the basic structure
provided by a regular network with $K=2$ neighbors.} \label{graf3}
\end{figure}

\begin{figure}
\includegraphics[width=7cm,height=7cm]{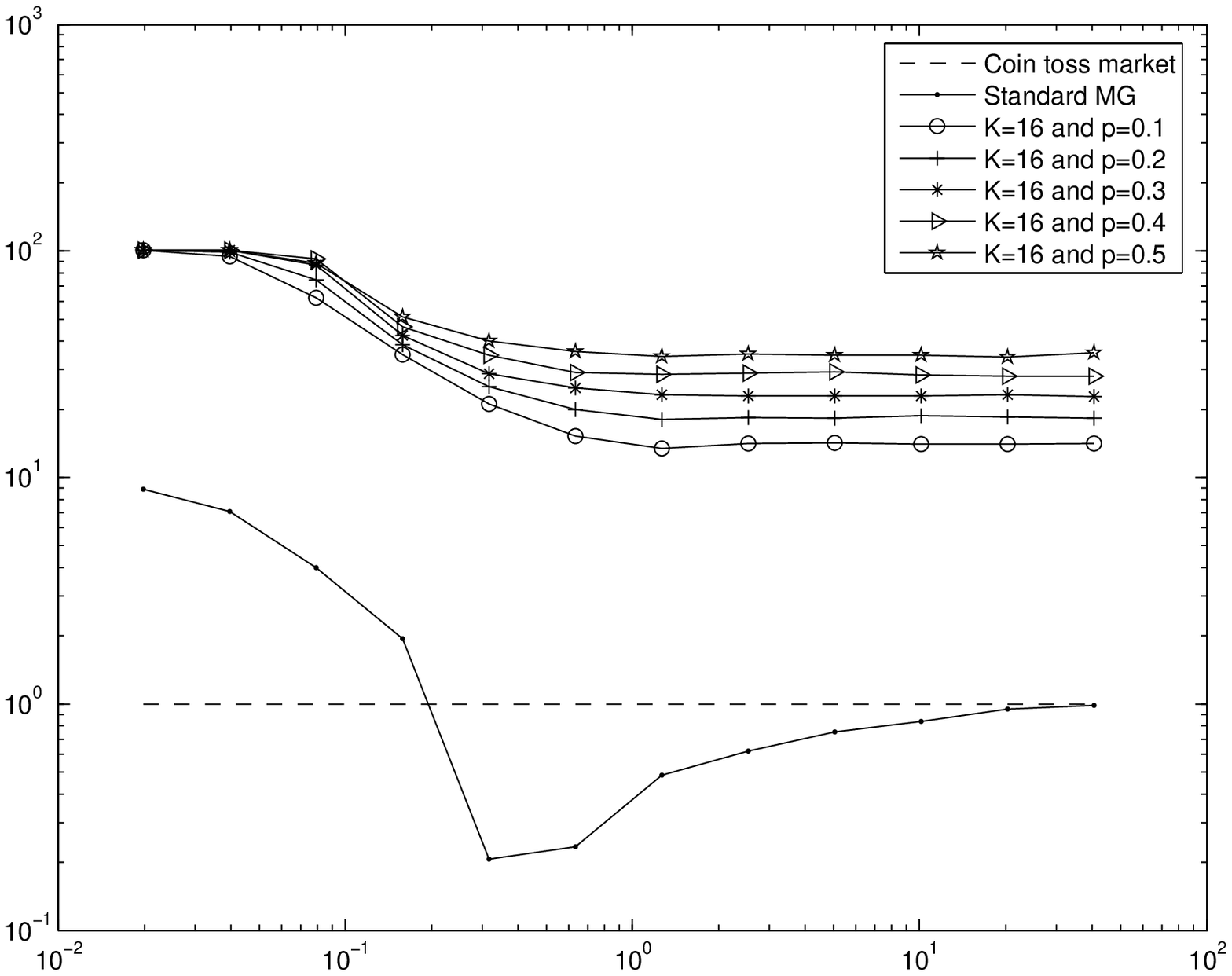}
\caption{The ratio $\sigma^2/N$ as a function of $\alpha=2^M/N$.
We compare here the standard MG with the modified MG presented in
this paper using small world networks with the basic structure
provided by a regular network with $K=16$ neighbors.}
\label{graf4}
\end{figure}

Figures \ref{graf1}, \ref{graf2}, \ref{graf3}
and \ref{graf4} present the main results of this paper. In all
figures, we have plotted $\sigma^2/N$ as a function of
$\alpha=2^M/N$ for the coin toss market, for the standard minority
game and for the minority game with the presence of herding
behavior. The difference among them is the structure of
neighborhood where in each figure the neighborhood is provided by
a different network. In these figures, all simulations used the
number of agents $N=101$, the number of strategies $S=2$ and the
time horizon $T=10000$. In figures \ref{graf1}, \ref{graf2} and
\ref{graf4}, $K$ is the number of ``regular'' neighbors -- a
parameter that arises in regular and small world networks. In
figures \ref{graf2}, \ref{graf3} and \ref{graf4}, $p$ is the
probability of two agents being connected -- a parameter that
arises in random graphs and small world networks.

First of all, one may notice that in the presence of herding
behavior, the volatility of the system is much larger than the
volatility of the standard MG. This happens because the presence
of herding behavior generates a crowd in the MG, i.e., a large
groups of agents using the same strategy. This fact has been
studied in economic theory~\cite{lee98} and
econophysics~\cite{sorjoh97}, which show that herding behavior may
be a source of large price movements and also crashes.

In all figures, up to a certain point of the presence of herding
behavior, the slope of the low-$M$ phase is not modified and this
phase is clear in almost all simulations. This can be easily
interpreted. In the original MG~\cite{savman99}, the low-$M$ phase
is characterized by a crowded phase where the number of strategies
is small when compared to the number of agents. The presence of
herding behavior only reinforces this fact. However, if the
presence of herding behavior is too strong as, for instance, in
figure \ref{graf4} where there are simulations with a small world
network with parameters $K=16$ and $p=0.5$, part of the standard
low-$M$ phase is replaced a horizontal line. This means that the
crowd is so strong that the additional number of strategies
introduced are not enough to reduce it.

On the other hand, in all figures, when the phenomenon of herding
behavior emerges, the high-$M$ phase is strongly affected. In
spite of the number of strategies is huge, this happens because
the number of strategies that the agents actually use is small
when compared to the number of agents.

Moreover, one should also notice that the value of $\alpha_c$ is
almost the same in all simulations when herding behavior is
present. However, the value of $\alpha_c$ in this situation is
larger than the value found in the standard MG.

\section{Final Remarks. }In this note, we have presented a new
version of the standard MG with local interactions that emerge due
to the presence of herding behavior. The herding behavior here
arises since the agents that play the game sometimes do not
believe in their own strategies and prefer to follow the most
informed agents that belong to their neighborhood. As it been
already pointed out, this kind of behavior is rational and
justified by the economic theory~\cite{ban92}.

Using this framework, we have found that the presence of herding
behavior may affect both the low-$M$ phase and the high-$M$ phase.
In particular, if one thinks the minority game as a model of
financial market\cite{chamar01}, then the results of this modified
model agrees with the results also found in the
economics~\cite{lee98} and econophysics~\cite{sorjoh97}
literature. Finally, we show that that the well known curve
volatility versus memory, which caracterizes the phases of a
minority game, is a monotone decreasing curve.

\clearpage

\bibliography{cajueiro_mghb}

\begin{thebibliography}{25}
\expandafter\ifx\csname natexlab\endcsname\relax\def\natexlab#1{#1}\fi
\expandafter\ifx\csname bibnamefont\endcsname\relax
  \def\bibnamefont#1{#1}\fi
\expandafter\ifx\csname bibfnamefont\endcsname\relax
  \def\bibfnamefont#1{#1}\fi
\expandafter\ifx\csname citenamefont\endcsname\relax
  \def\citenamefont#1{#1}\fi
\expandafter\ifx\csname url\endcsname\relax
  \def\url#1{\texttt{#1}}\fi
\expandafter\ifx\csname urlprefix\endcsname\relax\def\urlprefix{URL }\fi
\providecommand{\bibinfo}[2]{#2}
\providecommand{\eprint}[2][]{\url{#2}}

\bibitem[{\citenamefont{Challet and Zhang}(1997)}]{chazha97}
\bibinfo{author}{\bibfnamefont{D.}~\bibnamefont{Challet}} \bibnamefont{and}
  \bibinfo{author}{\bibfnamefont{Y.~C.} \bibnamefont{Zhang}},
  \bibinfo{journal}{Physica A} \textbf{\bibinfo{volume}{246}},
  \bibinfo{pages}{407} (\bibinfo{year}{1997}).

\bibitem[{\citenamefont{Arthur}(1994)}]{art94}
\bibinfo{author}{\bibfnamefont{W.~B.} \bibnamefont{Arthur}},
  \bibinfo{journal}{American Economic Review} \textbf{\bibinfo{volume}{84}},
  \bibinfo{pages}{406} (\bibinfo{year}{1994}).

\bibitem[{\citenamefont{Johnson et~al.}(2003)\citenamefont{Johnson, Jefferies,
  and Hui}}]{johjef03}
\bibinfo{author}{\bibfnamefont{N.~F.} \bibnamefont{Johnson}},
  \bibinfo{author}{\bibfnamefont{P.}~\bibnamefont{Jefferies}},
  \bibnamefont{and} \bibinfo{author}{\bibfnamefont{P.~M.} \bibnamefont{Hui}},
  \emph{\bibinfo{title}{Financial market complexity}}
  (\bibinfo{publisher}{Oxford University Press}, \bibinfo{address}{Oxford},
  \bibinfo{year}{2003}).

\bibitem[{\citenamefont{Coolen}(2005)}]{coo05}
\bibinfo{author}{\bibfnamefont{A.~C.~C.} \bibnamefont{Coolen}},
  \emph{\bibinfo{title}{The mathematical theory of minority games}}
  (\bibinfo{publisher}{Oxford University Press}, \bibinfo{address}{Oxford},
  \bibinfo{year}{2005}).

\bibitem[{\citenamefont{abd M.~Marsili and Zhang}(2005)}]{chamar05}
\bibinfo{author}{\bibfnamefont{D.~C.} \bibnamefont{abd M.~Marsili}}
  \bibnamefont{and} \bibinfo{author}{\bibfnamefont{Y.~C.} \bibnamefont{Zhang}},
  \emph{\bibinfo{title}{Minority games}} (\bibinfo{publisher}{Oxford University
  Press}, \bibinfo{address}{Oxford}, \bibinfo{year}{2005}).

\bibitem[{\citenamefont{Savit et~al.}(1999)\citenamefont{Savit, Manuca, and
  Riolo}}]{savman99}
\bibinfo{author}{\bibfnamefont{R.}~\bibnamefont{Savit}},
  \bibinfo{author}{\bibfnamefont{R.}~\bibnamefont{Manuca}}, \bibnamefont{and}
  \bibinfo{author}{\bibfnamefont{R.}~\bibnamefont{Riolo}},
  \bibinfo{journal}{Physical Review Letters} \textbf{\bibinfo{volume}{82}},
  \bibinfo{pages}{2203} (\bibinfo{year}{1999}).

\bibitem[{\citenamefont{Paczuski et~al.}(2000)\citenamefont{Paczuski, Bassler,
  and Corral}}]{pacbas00}
\bibinfo{author}{\bibfnamefont{M.}~\bibnamefont{Paczuski}},
  \bibinfo{author}{\bibfnamefont{K.~E.} \bibnamefont{Bassler}},
  \bibnamefont{and} \bibinfo{author}{\bibfnamefont{A.}~\bibnamefont{Corral}},
  \bibinfo{journal}{Physical Review Letters} \textbf{\bibinfo{volume}{84}},
  \bibinfo{pages}{3185} (\bibinfo{year}{2000}).

\bibitem[{\citenamefont{Kalinowski et~al.}(2000)\citenamefont{Kalinowski,
  Schulz, and Briese}}]{kalsch00}
\bibinfo{author}{\bibfnamefont{T.}~\bibnamefont{Kalinowski}},
  \bibinfo{author}{\bibfnamefont{H.-J.} \bibnamefont{Schulz}},
  \bibnamefont{and} \bibinfo{author}{\bibfnamefont{M.}~\bibnamefont{Briese}},
  \bibinfo{journal}{Physica A} \textbf{\bibinfo{volume}{277}},
  \bibinfo{pages}{502} (\bibinfo{year}{2000}).

\bibitem[{\citenamefont{Rios}(2002)}]{moerio02}
\bibinfo{author}{\bibfnamefont{S.~M. P. D.~L.} \bibnamefont{Rios}},
  \bibinfo{journal}{Physica A} \textbf{\bibinfo{volume}{303}},
  \bibinfo{pages}{217} (\bibinfo{year}{2002}).

\bibitem[{\citenamefont{Galstyan and Lerman}(2002)}]{galler02}
\bibinfo{author}{\bibfnamefont{A.}~\bibnamefont{Galstyan}} \bibnamefont{and}
  \bibinfo{author}{\bibfnamefont{K.}~\bibnamefont{Lerman}},
  \bibinfo{journal}{Physical Review E} \textbf{\bibinfo{volume}{66}},
  \bibinfo{pages}{015103} (\bibinfo{year}{2002}).

\bibitem[{\citenamefont{h.~J.~Quan et~al.}(2003)\citenamefont{h.~J.~Quan, Wang,
  Hui, and X.~S}}]{quawan03}
\bibinfo{author}{\bibnamefont{h.~J.~Quan}},
  \bibinfo{author}{\bibfnamefont{B.~H.} \bibnamefont{Wang}},
  \bibinfo{author}{\bibfnamefont{P.~M.} \bibnamefont{Hui}}, \bibnamefont{and}
  \bibinfo{author}{\bibfnamefont{L.}~\bibnamefont{X.~S}},
  \bibinfo{journal}{Physica A} \textbf{\bibinfo{volume}{321}},
  \bibinfo{pages}{300} (\bibinfo{year}{2003}).

\bibitem[{\citenamefont{Chau et~al.}(2004)\citenamefont{Chau, Chow, and
  Ho}}]{chacho04}
\bibinfo{author}{\bibfnamefont{H.~F.} \bibnamefont{Chau}},
  \bibinfo{author}{\bibfnamefont{F.~K.} \bibnamefont{Chow}}, \bibnamefont{and}
  \bibinfo{author}{\bibfnamefont{K.~H.} \bibnamefont{Ho}},
  \bibinfo{journal}{Physica A} \textbf{\bibinfo{volume}{332}},
  \bibinfo{pages}{483} (\bibinfo{year}{2004}).

\bibitem[{\citenamefont{Li and Savit}(2004)}]{lisav04}
\bibinfo{author}{\bibfnamefont{Y.}~\bibnamefont{Li}} \bibnamefont{and}
  \bibinfo{author}{\bibfnamefont{R.}~\bibnamefont{Savit}},
  \bibinfo{journal}{Physica A} \textbf{\bibinfo{volume}{335}},
  \bibinfo{pages}{217} (\bibinfo{year}{2004}).

\bibitem[{\citenamefont{Burgos et~al.}(2004)\citenamefont{Burgos, Ceva, and
  Perazzo}}]{burcev04}
\bibinfo{author}{\bibfnamefont{E.}~\bibnamefont{Burgos}},
  \bibinfo{author}{\bibfnamefont{H.}~\bibnamefont{Ceva}}, \bibnamefont{and}
  \bibinfo{author}{\bibfnamefont{R.~P.~J.} \bibnamefont{Perazzo}},
  \bibinfo{journal}{Physica A} \textbf{\bibinfo{volume}{337}},
  \bibinfo{pages}{635} (\bibinfo{year}{2004}).

\bibitem[{\citenamefont{Shang and Wang}(2005)}]{shawan05}
\bibinfo{author}{\bibfnamefont{L.}~\bibnamefont{Shang}} \bibnamefont{and}
  \bibinfo{author}{\bibfnamefont{X.~F.} \bibnamefont{Wang}},
  \bibinfo{journal}{Forthcoming in Physica A}  (\bibinfo{year}{2005}).

\bibitem[{\citenamefont{Caridi and Ceva}(2004)}]{carcev04}
\bibinfo{author}{\bibfnamefont{I.}~\bibnamefont{Caridi}} \bibnamefont{and}
  \bibinfo{author}{\bibfnamefont{H.}~\bibnamefont{Ceva}},
  \bibinfo{journal}{Physica A} \textbf{\bibinfo{volume}{339}},
  \bibinfo{pages}{574} (\bibinfo{year}{2004}).

\bibitem[{\citenamefont{Kirley}(2005)}]{kir05}
\bibinfo{author}{\bibfnamefont{M.}~\bibnamefont{Kirley}},
  \bibinfo{journal}{Forthcoming Physica A}  (\bibinfo{year}{2005}).

\bibitem[{\citenamefont{Erd\'{o}s and R\'{e}nyi}(1960)}]{erdren60}
\bibinfo{author}{\bibfnamefont{P.}~\bibnamefont{Erd\'{o}s}} \bibnamefont{and}
  \bibinfo{author}{\bibfnamefont{A.}~\bibnamefont{R\'{e}nyi}},
  \bibinfo{journal}{Bulletin of the International Statistical Institute}
  \textbf{\bibinfo{volume}{38}}, \bibinfo{pages}{343} (\bibinfo{year}{1960}).

\bibitem[{\citenamefont{Watts and Strogatz}(1998)}]{watstr98}
\bibinfo{author}{\bibfnamefont{D.~J.} \bibnamefont{Watts}} \bibnamefont{and}
  \bibinfo{author}{\bibfnamefont{S.~H.} \bibnamefont{Strogatz}},
  \bibinfo{journal}{Nature} \textbf{\bibinfo{volume}{393}},
  \bibinfo{pages}{440} (\bibinfo{year}{1998}).

\bibitem[{\citenamefont{Watts}(1999)}]{wat99}
\bibinfo{author}{\bibfnamefont{D.~J.} \bibnamefont{Watts}},
  \emph{\bibinfo{title}{Small worlds: the dynamics of networks between order
  and randomness}} (\bibinfo{publisher}{Princeton University Press},
  \bibinfo{address}{Princeton}, \bibinfo{year}{1999}).

\bibitem[{\citenamefont{Cajueiro}(2005)}]{caj05}
\bibinfo{author}{\bibfnamefont{D.~O.} \bibnamefont{Cajueiro}},
  \bibinfo{journal}{Physical Review E} \textbf{\bibinfo{volume}{72}},
  \bibinfo{pages}{047104} (\bibinfo{year}{2005}).

\bibitem[{\citenamefont{Lee}(1998)}]{lee98}
\bibinfo{author}{\bibfnamefont{I.~H.} \bibnamefont{Lee}},
  \bibinfo{journal}{Review of Economic Studies} \textbf{\bibinfo{volume}{65}},
  \bibinfo{pages}{395} (\bibinfo{year}{1998}).

\bibitem[{\citenamefont{Sornette and Johansen}(1997)}]{sorjoh97}
\bibinfo{author}{\bibfnamefont{D.}~\bibnamefont{Sornette}} \bibnamefont{and}
  \bibinfo{author}{\bibfnamefont{A.}~\bibnamefont{Johansen}},
  \bibinfo{journal}{Physica A} \textbf{\bibinfo{volume}{245}},
  \bibinfo{pages}{411} (\bibinfo{year}{1997}).

\bibitem[{\citenamefont{Banerjee}(1992)}]{ban92}
\bibinfo{author}{\bibfnamefont{A.}~\bibnamefont{Banerjee}},
  \bibinfo{journal}{Quartely Journal of Economics}
  \textbf{\bibinfo{volume}{107}}, \bibinfo{pages}{797} (\bibinfo{year}{1992}).

\bibitem[{\citenamefont{Challet et~al.}(2001)\citenamefont{Challet, Marsili,
  and Zhang}}]{chamar01}
\bibinfo{author}{\bibfnamefont{D.}~\bibnamefont{Challet}},
  \bibinfo{author}{\bibfnamefont{M.}~\bibnamefont{Marsili}}, \bibnamefont{and}
  \bibinfo{author}{\bibfnamefont{Y.~C.} \bibnamefont{Zhang}},
  \bibinfo{journal}{Physica A} \textbf{\bibinfo{volume}{299}},
  \bibinfo{pages}{228} (\bibinfo{year}{2001}).

\end{thebibliography}

\end{document}